# Selective Purcell enhancement of two closely linked zero-phonon transitions of a silicon carbide color center

David O. Bracher[1,2], Xingyu Zhang[1], and Evelyn L. Hu[1]

[1] John A. Paulson School of Engineering and Applied Sciences, Harvard University, Cambridge, MA 02138

[2] Department of Physics, Harvard University, Cambridge, MA 02138

Abstract: Point defects in silicon carbide are rapidly becoming a platform of great interest for single photon generation, quantum sensing, and quantum information science. Photonic crystal cavities (PCC) can serve as an efficient light-matter interface both to augment the defect emission and to aid in studying the defects' properties. In this work, we fabricate 1D nanobeam PCCs in 4H-silicon carbide with embedded silicon vacancy centers. These cavities are used to achieve Purcell enhancement of two closely spaced defect zero-phonon lines (ZPL). Enhancements of >80-fold are measured using multiple techniques. Additionally, the nature of the cavity coupling to the different ZPLs is examined.

**Introduction:**

Spin-active point defects in a variety of silicon carbide (SiC) polytypes have recently elicited a great deal of interest as the basis for solid-state single photon sources, nanoscale quantum sensing, and quantum information science[1–10]. Such color centers in SiC offer access to emission at a variety of wavelengths, ranging from visible (600-800 nm)[7] to near-IR (850-1300 nm)[1,2,6]. Furthermore, in many polytypes, such as 4H- and 6H-SiC, the diatomic lattice and the crystal structure lead to a wider variety of defect properties than, for example, in diamond. For instance, the silicon vacancy (SiV) center can occupy either hexagonal or cubic lattice sites, exhibiting differences in the zero phonon line (ZPL) emission wavelength and the zero field spin splitting, depending on which site hosts the vacancy[11–14]. Similar variation is seen in the neutral divacancy[1]. Thus, SiC defects yield a wide variety of properties that can be selected as desired and these defects can provide high sensitivity to subtle variations in their local environments.

Readout of such local information can be greatly augmented by coupling these point defects to high quality optical cavities which can enhance their optical emission properties, enabling large increases in the emission rate into their ZPLs[15–18]. A number of groups have fabricated photonic crystal cavities in SiC, achieving passively measured (i.e. without emitters in the cavity) quality factors, $Q$, at telecom frequencies of $10^3$-$10^4$ in 3C-SiC[19–21] and 6H-SiC and of ~$7\times10^4$ in amorphous SiC[22]. Other work has shown cavity-enhanced measurements of a 3C-SiC color center[23,24]. However, there have not yet been demonstrations of significant coupling to and enhancement of the ZPL emission from SiC point defects.

In this work, we report on such enhancement in high-$Q$, 1D nanobeam PCCs with embedded SiV centers in 4H-SiC. We selectively couple to two closely spaced ZPL transitions, and achieve enhancement of greater than ~80-fold, comparing the increase in the ZPL intensity on and off resonance with the cavity mode. The degree of enhancement is also confirmed by measurement of the spontaneous emission fraction into the cavity-enhanced ZPL and measurement of fluorescence lifetime decrease on resonance. In addition, we demonstrate the responsiveness of the cavity to temperature-induced changes in the ZPL populations as well as to the polarization properties of the two transitions. Therefore, the cavity-coupling work demonstrated here represents important progress in the field of SiC defect photonics, both paving the way towards the improved functionality of SiC defects for quantum sensing and information as well as establishing the role of the cavities themselves in providing valuable insights into the study of these defects.

**Results:**

**Characterization of PCCs and embedded point defects:**

As in our previous work, we employ a PCC design that provides a high theoretical quality factor, $Q$, with a small mode volume, $V$[25,26]. Such parameters are desirable given that the degree of emission enhancement provided by the cavity can be quantified by the Purcell factor[27],

$$F = \frac{3}{4\pi^2} \xi \frac{Q}{V} \left(\frac{\lambda}{n}\right)^3$$

The factor $\xi$ takes into account the spectral and spatial overlap of the cavity field with the emitter being enhanced, as well as the alignment of the cavity field and the emitter's optical dipole. With perfect overlap and alignment, $\xi$ is 1. Thus, for a given $\xi$, maximizing emission enhancement depends upon fabricating cavities with high $Q$ and low $V$. To provide high $Q/V$, the PCC design we have chosen consists of a linear array of holes (Fig. 1a), where the 8 innermost holes on either side of the center are linearly tapered to define a cavity region. The holes are tapered in both their center-to-center spacing and their parallel-to-beam diameter (x-axis in fig. 1b), creating elliptical holes in this inner region. For both parameters, the minimum tapered value is 84% of the values in the outer holes. FDTD simulations (Lumerical Solutions) confirm that these cavities have high theoretical $Q$ of 4 x $10^6$ and small mode volume of 0.45 $(\lambda/n)^3$, where $\lambda$ is the resonant wavelength of the cavity mode and $n$ the refractive index. The simulated electric field profile of the primary TE mode is shown Fig. 1b, showing that the fields are well localized to the inner, tapered cavity region.

The fabrication technique we employed, described in the Methods section, resulted in chips with ~3,000 cavities, designed with cavity resonances that span the full range of the SiV emission. The point defects were introduced after the fabrication of the cavities: the wafers with PCCs were ion implanted (Cutting Edge Ions Inc., see Methods) without post-annealing to create ensembles of SiV centers. Photoluminescence (PL) spectra of the implanted samples taken at 4K in a helium-cooled continuous flow cryostat (Janis) show the positions of the ZPLs of both the V1 and V2 types of the SiV center. These two centers correspond to the two inequivalent Si lattice sites in 4H-SiC, with V1 belonging to the cubic ($k$) site and V2 to the hexagonal ($h$) site[12]. The defects at both sites have total spin $S = 3/2$ and show ODMR signals[3,4,12]. As seen in Fig. 2, the V1 ZPL is observed at 861.4 nm, and the V2 ZPL at 916.5 nm, corresponding to values previously reported in literature. Moreover, the V1 center is known to have a high-temperature companion ZPL, typically denoted V1', stemming from a transition to an excited state of slightly higher energy than the excited state corresponding to the V1 transition[13]. Because of the 4.4 meV energy splitting between the V1' and V1 excited states (schematic shown in Fig. 2e), the intensities of the corresponding ZPLs are temperature dependent[13]. This is clearly observed in the spectra shown in Fig. 2 2: when the excitation laser is at a power of 100 μW, emission into the V1' line is suppressed (Fig. 2c, d). However, at a higher excitation laser power of 600 μW, the V1' transition intensity is nearly comparable to that of the V1 peak due to laser-induced heating (Fig. 2a, b). At the higher laser power, the branching ratio into each ZPL is ~1-1.3%, while at lower laser power the branching ratio into the V1 line is ~1.6% with only ~0.5% into the V1' line.

After ion implantation, we characterized the fabricated PCCs using confocal PL microscopy. A pulsed Ti:Sapphire laser tuned to a wavelength of 760 nm was used to optically excite the SiV centers in the cavities, and the emitted light was passed to a spectrometer for analysis. Measured PCC spectra showed cavity modes decorating the defects' broad PL signal (consisting primarily of phonon sideband emission). Even without post-implantation annealing of the material, the cavities displayed high values of $Q$, typically between ~2,000 and 5,000 (Fig. 1c), as calculated by fitting the cavity spectrum to a Lorentzian profile. Using $Q$ of 2,000 and the simulated $V$ of ~0.45 $(\lambda/n)^3$ in Eq. 1, we calculate a maximum Purcell enhancement of ~340, assuming $\xi$ of 1. Even without perfect defect placement within the cavity, significant ZPL enhancement can be achieved, as long as the overlap of the defect spatial location and cavity field maximum is sufficient to yield $\xi$ of ~0.1 or greater.

**Cavity tuning spectral measurements:**

To demonstrate such enhancement, it is easiest to choose devices with modes that are slightly blue-shifted from the desired ZPL wavelength and then slowly redshift the cavity mode through resonance. Therefore, we selected cavities with well-coupled modes that were blue-shifted relative to the 859.1 nm V1' line and used nitrogen gas condensation[18] to redshift the cavity mode and observe its tuning through both V1' and V1 ZPLs. By focusing on the interaction with these two ZPLs, we use the cavity as an exquisitely sensitive probe of these transitions which have the same ground state and two different excited states. The results of this tuning with one such cavity using an excitation power of 600 µW are shown in Fig. 3. An intensity map as a function of tuning step in the vicinity of the V1 and V1' ZPLs is shown in Fig. 3a. The excitation power and collection time were kept constant for each spectrum acquired so that the intensities can be directly compared. As labeled, the two fixed lines correspond to the V1 and V1' ZPLs, through which the cavity mode sweeps as it is red-shifted by the condensed gas. The intensity of each ZPL reaches a maximum when it coincides with the cavity mode wavelength.

Additionally, Figs. 3b-d provide details of the emission spectra at specific tuning steps to better underscore the increased emission into the ZPLs when they are resonant with the cavity mode. In Fig. 3b, we see the cavity mode far off resonance from both ZPLs. At cryogenic temperatures, the defects emit very little light at the cavity wavelength, and therefore emission into the ZPLs is of low intensity. Fitting the spectrum reveals a cavity $Q$ of ~2,200. The corresponding FWHM of the cavity, 0.4 nm, is well matched to the ZPL linewidths (on the order of ~0.5 nm), which aids in efficient coupling. In Fig. 3c and 3d, the cavity mode is fully on resonance with the V1' and V1 ZPLs, respectively. The large corresponding increases in intensity are readily apparent and we moreover see that the on-resonance intensity of the V1' is significantly higher than that of the V1. A typical means of quantifying the degree of Purcell enhancement achieved by a cavity is to compare the on- and off-resonant ZPL intensity[17,18]. Indeed, we have that

$$F = \frac{I_{ZPL,on}}{I_{ZPL,off}} - 1.$$

From this equation, we obtain an enhancement of 75 for the V1' ZPL and 22 for the V1 ZPL in the cavity described in Fig. 3.

We then carried out the same cavity tuning but using a lower excitation power of 100 µW. These data are shown in Fig. 4, with Figs. 4b-d showing the detailed spectra of the cavity mode far off resonance and then on resonance with the V1 and V1' ZPLs. Using on- and off-resonance intensity measurements, we derive the same degree of enhancement as observed for the higher-power (600 µW) excitation: an enhancement of 75 for the V1' ZPL and 22 for the V1 ZPL. However, while the cavity enhancement for both transitions is thus shown to be independent of excitation power used, the ratio of the on-resonance intensities for the V1' vs. V1 ZPLs is much larger in the 600 µW (Fig. 3c, d) case than in the 100 µW case (Fig. 4c, d). This arises from the fact that the relative intensity ratio of the V1'/V1 ZPLs at 600 µW (~1.3x), compared to at 100 µW excitation (~3x), is the same whether on or off resonance. Thus, the cavity is quite specific in its enhancement of these two closely-spaced and related transitions. Furthermore, by controlling both the excitation power and the ZPL to which the cavity is coupled, maximum contrast of emission into one ZPL or the other as desired can be achieved.

**Measurements of spontaneous emission enhancement:**

To confirm that we are achieving the spontaneous emission enhancement that is characteristic of the Purcell effect, we present further measurements on a second cavity, which showed the largest measured Purcell enhancement. We begin by considering the spontaneous emission factor, $\beta$, which can be calculated as the fraction of light emitted into the ZPL with the cavity mode on resonance, $I_{ZPL,on}/I_{total}$[28]. We can relate $\beta$ to the Purcell factor by the equation[29]

$$\beta = \frac{F\Gamma}{1 + (F-1)}$$

where $\Gamma$ is the branching ratio into the ZPL when the cavity is off resonance. This can be rearranged to give

$$F = \frac{\beta(1-\Gamma)}{\Gamma(1-\beta)}$$

Fig 5a shows the full PL spectra of this cavity. The upper and lower spectrum show the cavity on resonance with the V1' and V1 ZPLs, respectively. The inset shows the off-resonance spectrum. In this cavity, for the V1' (V1) transitions, the values of $\beta$ are 47.4% (23.7%), and the values of $\Gamma$ are 1% (1.2%). These correspond to $F \sim 89$ and $\sim 25$ for the V1' and V1 ZPLs, respectively, which match the values derived by comparing the total intensity of the ZPLs on and off resonance, ~90 and ~26. Furthermore, these measurements demonstrate that the ZPL is being selectively enhanced without a corresponding increase in the sideband emission, as desired. Therefore, using these cavities, we are able to direct nearly half of the SiV emission into the V1' ZPL, as opposed to the natural branching ratio of ~1%. Achieving such an increase is extremely important for quantum information applications using these point defects (e.g. entanglement) that can only make use of ZPL emission[15–17].

We further corroborate the spontaneous emission enhancement by measuring the excited state lifetime of the SiV defect on resonance with the cavity. The Purcell enhancement can be calculated from lifetime measurements using the following relation[16,28,29]:

$$F = \frac{\tau_0}{\Gamma}\left(\frac{1}{\tau_{on}} - \frac{1}{\tau_{off}}\right)$$

Here, $\tau_{on}$ refers to the lifetime when the mode is on resonance with the ZPL, $\tau_{off}$ the lifetime when the mode is off resonance, and $\tau_0$ the natural lifetime in the bulk material. When on resonance, the fluorescence lifetime decreases from $\tau_0$ due to the increased photonic density of states. When the cavity analyzed in Fig. 5a was tuned into resonance with the V1' ZPL, the ZPL emission was selected using a narrow bandpass filter (860 nm center wavelength, 10 nm FWHM) and passed to an avalanche photodiode for single photon counting. Fig. 5b shows the resultant fluorescence decay signal (green curve), which is fit to yield an on-resonance lifetime of 4.2 ns. We also measure an off-resonance lifetime of 9.4 ns (blue curve) and a bulk lifetime of 6.4 ns (red curve). The bulk lifetime matches previous measurements of the silicon vacancy fluorescence lifetime[6]. Moreover, as expected, the off-resonance lifetime is longer than the bulk lifetime due to a reduced density of states when the cavity mode is off resonance with the ZPL. Using these values yields an estimated Purcell enhancement of 84, which substantiates the value estimated from measurements of ZPL intensity enhancement and increased spontaneous emission fraction into the ZPL.

**Discussion and Conclusion:**

Lastly, we further explore the different enhancement factors for the V1' ZPL as compared to the V1 ZPL. The ~3.5-fold greater enhancement of the V1' ZPL in both cavities presented in this work is also recapitulated in measurements of all other cavities tuned through resonance. Since we are probing ZPLs that are very close spectrally, the $Q$ and $V$ of a given cavity change negligibly as it is tuned through resonance with both ZPLs. Moreover, since the ZPLs result from transitions of the same SiV centers (those sitting at the $k$ lattice sites within the cavity volume), the emitter-field spatial overlap of the ensemble of defects being enhanced in a given cavity does not vary for the two ZPLs. Therefore, the most likely explanation for the difference in enhancement between the V1' and V1 ZPLs is a difference in alignment between the cavity field vector and the emitter optical dipole. Our SiC crystal is grown along the *c*-axis, and therefore the plane of the PCCs (and correspondingly the cavity mode polarization) is perpendicular to the *c*-axis. Previous work has suggested that the V1' transition is strongly polarized perpendicular to the *c*-axis and the V1 transition is weakly polarized parallel to the *c*-axis[13]. Such polarization would explain well the difference in enhancement between the two ZPLs. Indeed, if the optical dipole for the V1' transition is assumed to be nearly perpendicular to the *c*-axis (and thus closely aligned to the cavity mode), the values of enhancement measured here would imply that the V1 optical dipole is polarized approximately 60 degrees from the V1' dipole, giving an angle of ~30 degrees with the c-axis.

In summary, we have demonstrated 1D nanobeam PCCs in 4H-SiC with high-quality factor modes ($Q$ up to 5,300) coupled to silicon vacancy defect centers generated by ion implantation. Using high-$Q$ cavities with resonances slightly blue-shifted from two closely related silicon vacancy ZPLs, we used nitrogen gas condensation to controllably tune the cavity mode into resonance with these ZPLs. We observed Purcell enhancement of more than 80-fold and confirm these measurements by comparing on- and off-resonant intensities, increased spontaneous emission fraction into the ZPLs, and decreased defect fluorescence lifetime on resonance. We were able to direct up to nearly half the light emitted by the SiV into the V1' ZPL, a substantial improvement over the natural branching fraction. Additionally, the responsiveness of the cavity to changes in ZPL population and to differences in polarization shows that the cavity can serve as effective and highly sensitive probe to both amplify and study the optical properties of the defects. For example, it has been noted elsewhere that the optical signatures of the V1, V1' and V2 defects are sensitive to strain in the material[13]. These cavities could therefore provide precise and dynamic assessments of the local strain environments of defects at different characteristic sites in the SiC lattice. Thus, with the demonstrated ability to achieve a high degree of cavity-defect interaction in silicon vacancy centers in 4H-SiC, this work represents an important step in the development of point defects in SiC as a significant platform for solid state quantum devices.

**Methods:**

**Sample fabrication:** Electron-beam lithography (Elionix) is used to define the cavity pattern, which is transferred to an aluminum hard mask by reactive ion etching (Unaxis Shuttleline). A second RIE step (STS ICP-RIE) transfers the pattern to the SiC wafer, which consists of a 250 nm p-type epilayer homoepitaxially grown on an n-type substrate. After pattern transfer, the n-type substrate is removed with a dopant-selective photoelectrochemical etch to provide optical isolation for the cavity[25,30]. This procedure allows us to create cavities out of high-quality epitaxially grown material.

**Defect implantation:** Silicon vacancy centers were created by implantation of $C_{12}$ ions at a dose of $10^{12}$ ions/cm$^2$ and an energy of 70 keV. In the vertical dimension, the cavity field profile reaches its maximum value in the middle of the cavity, which is 125 nm below the surface for our 250-nm device layer. Therefore, the 70 keV ion energy was selected to create defects with a maximum probability at this depth, as calculated by stopping range of ions in matter (SRIM) simulations.

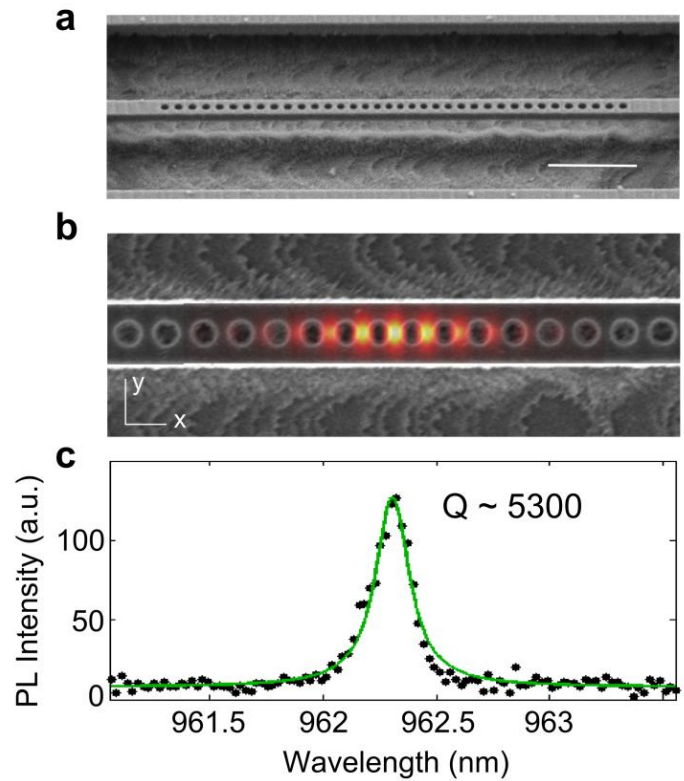

Figure 1. Nanobeam photonic crystal cavity (PCC) and mode. (a) SEM micrograph of 1D nanobeam PCC with embedded silicon vacancy (SiV) centers. Scale bar indicates 2 μm. (b) Detailed view of inner cavity region of the PCC with overlaid electric field profile from FDTD simulations. (c) Measured cavity mode coupled to SiV luminescence showing high quality factor, *Q* of 5,300. Lorentzian fit to measured data is shown in green.

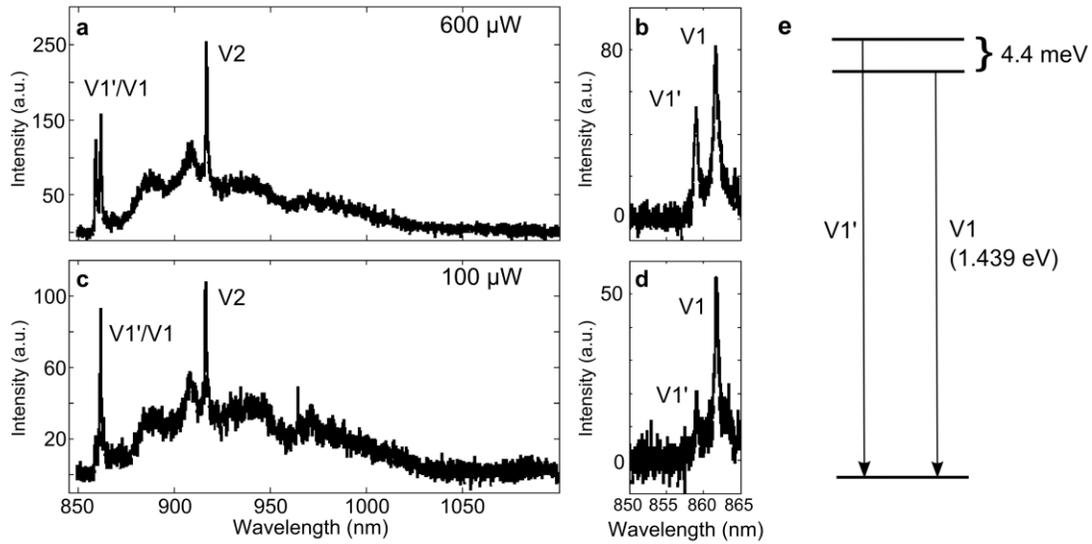

Figure 2. SiV spectrum and ZPL intensity. (a) Full spectrum of SiV luminescence measured with 600 μW excitation power at 4K. The V1', V1, and V2 ZPLs are all visible and indicated. (b) High-resolution spectrum showing V1' and V1 ZPLs measured with 600 μW excitation power at 4K. (c),(d) Same as (a),(b) with 100 μW excitation power. Here, the V1', whose intensity is temperature-dependent, has much lower intensity due to much reduced laser-induced heating. (e) Schematic showing energy level structure corresponding to V1 and V1' ZPL emission.

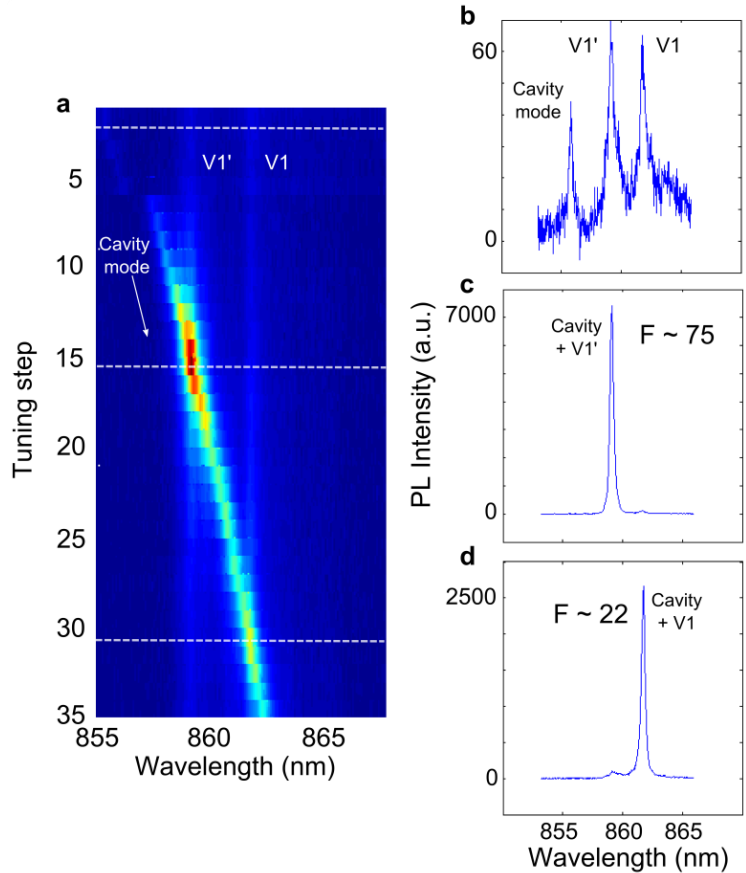

Figure 3. Cavity mode tuning and ZPL enhancement at 600 μW excitation (a) Intensity map showing the interaction of the cavity mode with the V1' and V1 ZPLs as a function of nitrogen gas injection step. As the gas is injected, the cavity mode redshifts (as indicated), while the V1' and V1 ZPLs remain fixed. For visibility, the square root of the intensity is plotted. The three dotted white lines correspond to cross sections shown in (b)-(d). (b) Spectrum taken when the cavity mode is far off-resonance, showing the cavity mode and the two ZPLs. (c),(d) Spectra taken when the cavity is resonant with the V1' and V1 ZPLs, respectively. The marked increase in intensity on resonance can be clearly seen. As indicated, the V1' line is enhanced ~75-fold, and the V1 line is enhanced ~22-fold on resonance.

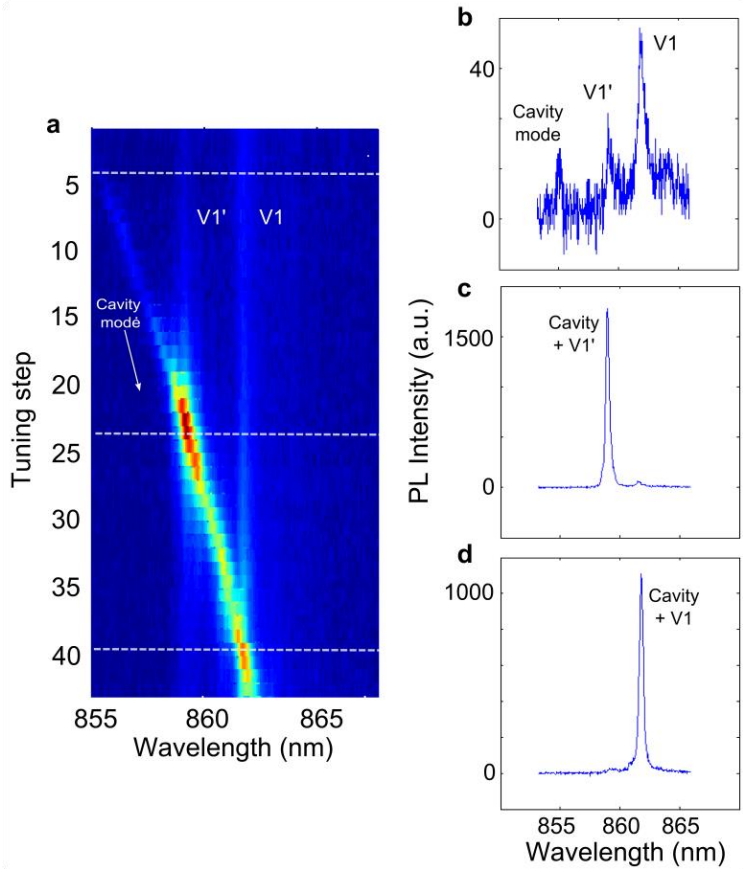

Figure 4. Cavity mode tuning and ZPL enhancement at 100 µW excitation (a) Intensity map showing the interaction of the cavity mode with the V1' and V1 ZPLs as a function of gas injection step. As in Fig. 3, the redshift of the cavity mode is indicated, and the three dotted white lines correspond to cross sections shown in (b)-(d). (b) Spectrum taken when the cavity mode is far off-resonance showing the cavity mode and the two ZPLs. (c),(d) Spectra taken when the cavity is resonant with the V1' and V1 ZPLs. As compared to Fig. 3, the relative intensities of the ZPLs on resonance are much closer, because the V1' line, while exhibiting a higher degree of enhancement, has much lower intensity than the V1 line at this excitation power. As in Fig. 3, the ZPL intensities are enhanced by factors of ~75 and ~22.

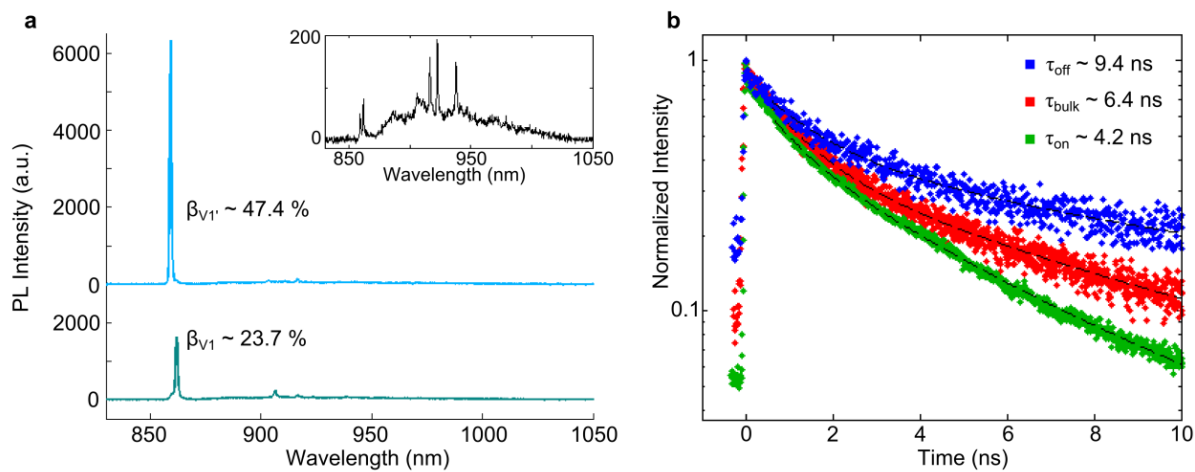

Figure 5. Increase in ZPL spontaneous emission rates and reduction in fluorescence lifetime. (a) Full spectrum of cavity showing highest Purcell enhancement when on resonance with V1' ZPL (upper) and V1 ZPL (lower), labeled by the fraction of light emitted into the ZPLs on resonance. The inset shows the spectrum taken when the cavity is blue-shifted from both ZPLs. Nearly 50% of the SiV emission is directed into the V1' ZPL on resonance. (b) Fluorescence lifetime measurements when the cavity mode is blue-shifted from the ZPLs (blue) and on resonance with the V1' ZPL (green). A lifetime measurement of the SiV in bulk (red) is also shown. As expected, the emitter lifetime is increased when the cavity mode is off resonance and decreased when it is on resonance.

**Acknowledgements:** The authors acknowledge funding from the Air Force Office of Scientific Research, under the QUMPASS program (Award No. FA9550-12-1-0004). This work was performed in part at the Center for Nanoscale Systems (CNS), a member of the National Nanotechnology Infrastructure Network (NNIN), which is supported by the National Science Foundation under NSF award no. ECS-0335765. CNS is part of Harvard University.

**Author Contributions:** D.O.B. optimized, fabricated, and characterized the cavities; performed cavity tuning measurements; analyzed the data; and prepared the manuscript. X.Z. helped make the cavity tuning measurements. E.L.H. conceived and directed the project. All authors discussed results and commented on the manuscript.

**Competing Interests:** The authors declare no competing financial interests.